\documentclass{article}
\usepackage[numbers]{natbib} 
\usepackage{amsmath}
\usepackage{float}
\usepackage{PRIMEarxiv}
\usepackage[table]{xcolor}
\usepackage[utf8]{inputenc} 
\usepackage[T1]{fontenc}    
\usepackage{hyperref}       
\usepackage{url}            
\usepackage{booktabs}       
\usepackage{amsfonts}       
\usepackage{nicefrac}       
\usepackage{microtype}      
\usepackage{lipsum}
\usepackage{fancyhdr}       
\usepackage{graphicx}       
\graphicspath{{media/}}     

\title{Web Archives Metadata Generation with GPT-4o: Challenges and Insights}

\author{
    Ashwin Nair \\
  Resource Discovery Management \\
  National Library Board, Singapore \\
  \texttt{ashwin\_nair\_madhavan@nlb.gov.sg} \\
   \And
  Goh Zhen Rong \\
   Nanyang Technological University, Singapore \\
  \texttt{zgoh033@e.ntu.edu.sg} 
   \And
Liu Tianrui \\
  National University of Singapore \\
  \texttt{tianrui\_liu@u.nus.edu} \\
  \And
    Abigail Yongping Huang \\
  Resource Discovery Management \\
  National Library Board, Singapore \\
  \texttt{abigail\_huang@nlb.gov.sg} \\
  }

\begin{document}
\maketitle

\begin{center}
    \text{All authors contributed equally to this work.}
\end{center}

\vspace{1em}

\begin{abstract}
Current metadata creation for web archives is time consuming and costly due to reliance on human effort. This paper explores the use of GPT-4o for metadata generation within the Web Archive Singapore, focusing on scalability, efficiency, and cost effectiveness. We processed 112 Web ARChive (WARC) files using data reduction techniques, achieving a notable 99.9\% reduction in metadata generation costs. By prompt engineering, we generated titles and abstracts, which were evaluated both intrinsically using Levenshtein distance and BERTScore, and extrinsically with human cataloguers using McNemar’s test. Results indicate that while our method offers significant cost savings and efficiency gains, human curated metadata maintains an edge in quality. The study identifies key challenges including content inaccuracies, hallucinations, and translation issues, suggesting that large language models (LLMs) should serve as complements rather than replacements for human cataloguers. Future work will focus on refining prompts, improving content filtering, and addressing privacy concerns through experimentation with smaller models. This research advances the integration of LLMs in web archiving, offering valuable insights into their current capabilities and outlining directions for future enhancements. The code is available at \url{https://github.com/masamune-prog/warc2summary} for further development and use by institutions facing similar challenges.\end{abstract}


\section{Introduction}
The digital landscape is constantly evolving, and the need to preserve our online heritage has become increasingly urgent. The Resource Discovery (RD) department of the National Library Board Singapore (NLB) provides cataloguing services for the collections of the National Library (NL) and public libraries in Singapore. NL, like many other archives and libraries worldwide, collects and archives websites in its effort to preserve the mercurial history of the web \cite{nlb}. This collection is known as Web Archive Singapore~\cite{NLB_Web_Archive_FAQ}. Each website is manually reviewed and catalogued according to the local application profile based on Dublin Core~\cite{dclibraries2004}. NL has been crawling websites on a curated basis since 2006, with each website requiring individual website owner consent. In 2019, the National Library Board Act was amended to impart NL with the authority to harvest all websites in the .sg domain without explicit consent from each individual website owner~\cite{NationalLibraryBoardAct1995}. The legislative change resulted in the explosive growth of the web collection~\cite{tay2020archive}. As of June 2024, there are over 98,000 unique websites and webpages from both the domain crawl and curated collection~\cite{NLB_Web_Archive_FAQ}. With the large collection numbers, RD sought to explore whether technology such as generative AI would be viable to facilitate the cataloguing of individual sites harvested in the yearly domain crawl. RD wished to explore the generation of titles, abstracts, and subjects for each website, as these metadata fields were deemed the most crucial and used for display and discovery.

\subsection{Background and Motivation}
This paper addresses the critical need for an efficient and accurate method of generating metadata for web archive collections. We focus on the challenge of managing large-scale web archives, where manual metadata curation is no longer practical due to resource constraints and the sheer volume of data.

\subsection{Problem Statement}
This paper addresses two primary challenges:
\begin{enumerate}
    \item \textbf{Efficiency:} How can we develop an automated system that can process and generate metadata for a large number of websites, significantly reducing the time and resources required?
    \item \textbf{Accuracy:} How can we ensure that the automatically generated metadata maintains a decent level of quality, accurately representing the content of the archived web pages?
\end{enumerate}
These challenges are nontrivial for collections such as WAS, which aims to preserve the digital heritage of Singaporean life, culture, and history in the 21st century. With an estimated 20,000 Web ARChive (WARC) files created annually that comply with quality control standards, there is a critical need for a scalable, reliable, and cost-effective solution for metadata generation.

\section{Related Work}

WARC files are a standardized format crucial for web crawling, archiving, and digital preservation~\cite{warc_format}.They play a significant role in historical web content preservation and the development of large language models (LLMs)~\cite{common_crawl}. For instance, Common Crawl, a major source of training data for LLMs, provides monthly crawls of billions of web pages in WARC format, significantly contributing to models such as GPT-3~\cite{brown2020language}. Additionally, WARC files ensure that historical web data remains accessible and support research into web trends and digital culture by offering detailed snapshots of web pages~\cite{warc_format}. This capability is essential for tracking changes and innovations, while the format’s design ensures effective accessibility and interoperability for both research and archiving~\cite{iipc_warc,sage_warc}.

\subsection{Metadata Strategies for Web Archives}
This work focuses on the provision of descriptive metadata to support search and access of individual sites in a web archive. There can be varying metadata approaches such as ~\cite{wong2024}:
\begin{itemize}
    \item harvesting the metadata as-is;
    \item reviewing only basic metadata fields such as 'Title' and 'Language', or
    \item providing fuller metadata for specially curated web collections, including assigning subject headings \cite{wong2024}
\end{itemize}
In many cases, a combination of approaches is used. For institutions conducting large-scale automated crawling (e.g., by domain), this would be:
\begin{itemize}
    \item for websites that have been curated into themes, metadata is individually reviewed and enhanced;
    \item for the remainder of the full crawl, metadata is used as-is; 
    \item for both, full-text search is available. 
\end{itemize}

In addition, there are also available metadata guidelines for web archiving such as those developed by OCLC Research’s Web Archiving Metadata Working Group~\cite{OCLC2024}.
In NLB, it was deemed viable for each site (but not subpages) in the domain crawl to be catalogued individually. This is implemented via a local application profile based on Dublin Core. Many fields are populated at the collection level, but cataloguers individually review the following fields which are used for searching and browsing (alongside full-text search): This is implemented via a local application profile based on Dublin Core. Many fields are populated at the collection level, but cataloguers individually review the following fields which are used for searching and browsing (alongside full-text search):
\begin{itemize}
    \item Title
    \item Abstract
    \item Subject (Library of Congress Subject Headings and a local categorization vocabulary of 17 top level terms)
\end{itemize}

Cataloguing every single website in the domain crawl is a time-consuming task and given the explosive progress and availability of artificial intelligence/machine learning (AI/ML) tools in recent years, especially with large language models (LLMs), we were keen to explore integrating this technology into our workflow.

Libraries have naturally explored myriad use cases for AI/ML. In the case of cataloguing, there have been many attempts at using generative AI for cataloguing and metadata enhancement~\cite{Brzustowicz2023,Chow2024}.The IFLA WLIC 2023 programme featured an entire session on “Utopia, Threat or Opportunity First? Artificial Intelligence and Machine Learning for Cataloguing,” with speakers from four different countries sharing on the use of AI/ML in cataloguing applications, such as generating catalogue records, detecting metadata from scanned resources, and deduplication~\cite{IFLA2023}.

While results have been promising, most projects embody what was said by Allen in a Library of Congress blog post:
\begin{quote}
"With all the possibilities, why aren’t we already embedding machine learning into everything we do? The answer is complicated, but at its simplest, machine learning (ML) and artificial intelligence (AI) tools haven’t demonstrated that they’re able to meet our very high standards for responsible stewardship of information in most cases, without significant human intervention"\cite{Allen2024}.\end{quote} 

Interest in AI/ML has naturally extended to the web archive community. This was embodied in the program for the 2024 IIPC Web Archiving Conference, which featured a dedicated session themed “AI \& Machine Learning,” in addition to having other lightning and drop-in talks that featured AI/ML use~\cite{WAC2024}.  We know of many projects utilizing AI/ML for web archives such as the Harvard Law School Library Innovation Lab’s WARC-GPT, launched in February 2024, which is a custom chatbot for users to explore WARC collections via natural language queries~\cite{Cargnelutti2024}.  To date, however, we have not come across specific work on the application of AI/ML for cataloguing by directly processing WARC files.

\subsection{Large Language Models}
The advent of LLMs has revolutionized the field of natural language processing (NLP). These sophisticated models, capable of processing and generating human-like text, have surpassed traditional methods in numerous NLP tasks. What was once considered a complex challenge, such as text summarization, is now routinely managed with impressive accuracy and efficiency by LLMs. Leading the charge in the commercialization of large language models are models such as GPT-4o by OpenAI, Claude by Anthropic, and Gemini by Google~\cite{gpt4o,claude,gemini}.  Built upon the groundbreaking transformer architecture introduced in “Attention Is All You Need,” these models leverage the attention mechanism to outperform earlier methods for NLP tasks that relied heavily on linguistic techniques like stemming and lemmatization or other neural networks like recurrent neural networks (RNN) and gated recurrent units (GRU)~\cite{attention2017need}.  The integration of reinforcement learning from human feedback  and direct preference optimization  further enhances the quality of LLM-generated output, making them more human-like and thus suitable for generation of high-quality content~\cite{rlhf,dpo}. Fine-tuning LLMs with datasets from a particular domain not only boosts their contextual understanding and expertise in domain specific tasks but also enhances the accuracy of their output~\cite{wu2023pmcllamabuildingopensourcelanguage,liu2023fingptdemocratizinginternetscaledata}.

\subsection{In-Context Learning}
However, given the limitations of time and computational resources, the process of fine-tuning language models for specific downstream tasks often incurs significant costs. As a result of this challenge, in-context learning (ICL) has gained significant popularity. The ICL technique enables LLMs to acquire new tasks without modifying the model parameters, by relying on task-specific examples provided in the input context. ICL has demonstrated strong performance in few-shot and zero-shot learning. Zero-shot learning leverages the model’s existing knowledge to deduce task requirements. As illustrated by Brown et al., GPT-3 was utilized to showcase the system’s versatility in the absence of task-specific fine-tuning. With few-shot learning, the model is provided with a restricted number of examples to showcase the task, resulting in enhanced comprehension and performance. Radford et al. have demonstrated that few-shot prompting can significantly enhance the model’s accuracy across a diverse range of tasks\cite{radford2021learning}.

Multi-query reasoning techniques and single-query reasoning techniques like chain of thought (CoT) prompting are examples of advanced ICL techniques. CoT prompting directs the model to produce intermediate reasoning steps that resemble how people solve problems. This approach is efficient and effective because it guarantees acceptable outcomes in addition to lowering the token count. The use of CoT significantly improves performance on arithmetic and logic-based activities~\cite{wei2022chain}. Limited in number of instances, prompting is a supplementary method that involves providing task-specific examples to assist in generating answers. 

Multi-query reasoning techniques, like graph of thought (GoT), tree of thought (ToT), and least-to most, use multiple LLM queries to extract various tenable reasoning paths~\cite{besta2024graph,yao2023tree,zhou2023least}.  By breaking down complicated issues into smaller, more manageable subproblems, these techniques improve the reasoning power of the model. In conclusion, in-context learning strategies are essential for fully utilizing LLMs. The rationale behind us selecting CoT is its capacity to minimize token consumption while yielding superior outcomes.
\section{Methodology}
\subsection{Data Collection and Preparation}
To obtain web data for our inquiry, we collected a total of 112 WARC (Web ARChive) files from the Web Archives of Singapore. The HTML content from these files was extracted using the WARCIO and fastwarc python libraries~\cite{warcio,bevendorff2021fastwarc}. BeautifulSoup facilitated the extraction of relevant metadata, such as titles and primary text content, from the HTML. We removed unnecessary tags and scripts during this procedure to ensure that the main content is highlighted. We standardized the URLs and conducted a quality assurance assessment to eliminate any substandard or irrelevant data, ensuring uniformity. This involved identifying common indicators of nonfunctional material, such as “404” errors or placeholder text like “lorem ipsum.” In addition, we implemented a deduplication technique to consolidate individual records obtained from duplicate URLs. This ensured the preservation of the information’s originality and relevance.
We employed multithreaded processing using a ThreadPoolExecutor to efficiently handle the vast amount of data, resulting in optimized resource utilization and significantly reduced processing time. The final step was combining the processed data into DataFrames, which were then ready for further analysis and model development. An indispensable requirement for the reliability and robustness of our investigation was a high caliber dataset ensured by this approach.
The final step was combining the processed data into DataFrames, which were then ready for further analysis and model development. An indispensable requirement for the reliability and robustness of our investigation was a high-caliber dataset ensured by this approach.

\subsection{Heuristics for Data Reduction}
To reduce the number of input tokens and thereby achieve cost savings, our methodology involved developing and evaluating various heuristic methods for efficient content extraction and summarization. The three heuristics were: 
\begin{enumerate}
\item \textbf{About Page Priority:} We prioritized extracting content from the “About” page. If no “About” page existed, we used content from the shortest URL.
\item \textbf{Shortest URL:} We extracted content exclusively from the web page with the shortest URL.
\item \textbf{Shortest URL with Regex Filtering:} We extracted content from the shortest URL and applied regular expression (regex) filters to reduce the token count, thereby optimizing the input for our model.\end{enumerate}These heuristics were inspired by observations of professional cataloguers, who typically require only a few pages to make accurate judgments about content categorization. This step reduces the token count across all heuristic methods. By adopting this methodology, we systematically compared various strategies for content extraction and summarization, striking a balance between computational efficiency and accuracy in content representation.

\subsection{Prompt Engineering for Title and Abstract Generation }
After some initial testing with in-context learning in LLMs, we settled on two meticulously designed prompts with contrasting characteristics for generating titles and abstracts to enhance metadata accuracy and website classification. These CoT prompts guide the generation process by providing clear instructions for cataloguing and summarizing website content. The output was then subjected to both automated and manual evaluation processes. This prompt, which builds upon the previous one, provides specific rules for the summarization of a variety of websites, such as corporate websites, personal blogs, and property listings. It guarantees that the summary is appropriate for the website’s type, thereby enhancing the relevance and precision of the abstract that is generated.
\\\\Prompt 1: \\
{\itshape
You are a diligent cataloguer working to create metadata for websites. Let's think step by step to ensure accurate and comprehensive metadata creation:

1) Determine the title of the organization or company on the main web page, ensuring it reflects the primary focus or name without additional descriptors. It should match the root domain of the web page.

2) Create an abstract: Summarize the main content of the website in a brief and informative abstract.

3) Format the Result: Return the result in JSON format as \{'title': [inferred\_title], 'abstract': [created\_abstract]\}.

This prompt initiates the process by identifying the main title and summarizing the website's content. The emphasis is on creating a concise and informative abstract and ensuring the title aligns with the website's root domain. The result is formatted in JSON to facilitate structured data use.
}
\\\\Prompt 2 (appended to Prompt 1):

{\itshape Summarize the content of the website following these rules:

- For company websites:
This is the website of (company’s name) which offers (services). The website contains information of (contact, operating hours, location, its services, customers’ testimonials).

- For websites selling properties:
(Name of project) is a private residential development by (name of company). The project is located at xxx. This website contains information on (the condominium, location, floor plans, developer and contact details).

- For personal websites/blogs:
This is a website of (Name of person), (role). This website contains information on (work experience, profile, education, research works, projects, publications, professional development, skills, portfolio).

- For others, create a summary.}

This prompt, which builds upon the previous one, provides specific guidelines for the compilation of cataloguing rules for the summarization of a variety of websites, such as corporate websites, personal blogs, and property listings. It guarantees that the summary is appropriate for the website’s type, thereby enhancing the relevance and precision of the abstract that is generated. 
These prompts are intended to improve metadata generation by directing the LLM to perform comprehensive, rule-based actions in a methodical manner. In the subsequent segment, we also demonstrate the contrast between the rule-based prompt result and the rule-free outcome. This ensures the development of abstracts and titles that are contextually pertinent, concise, and precise, which are critical for enhancing the searchability and classification of websites.

\subsection{Evaluation Methods}
\textbf{Automated Evaluation}: 

We employed an aggregation of two metrics to assess the quality of approaches used.
\begin{itemize}
    \item Levenshtein distance for title comparison ~\cite{levenshtein}
    \item BERTScore (using bert-based-cased as the embedding) for title and content similarity ~\cite{bertscore}
\end{itemize}

\begin{figure}[ht]
    \centering
    \includegraphics[width=\linewidth]{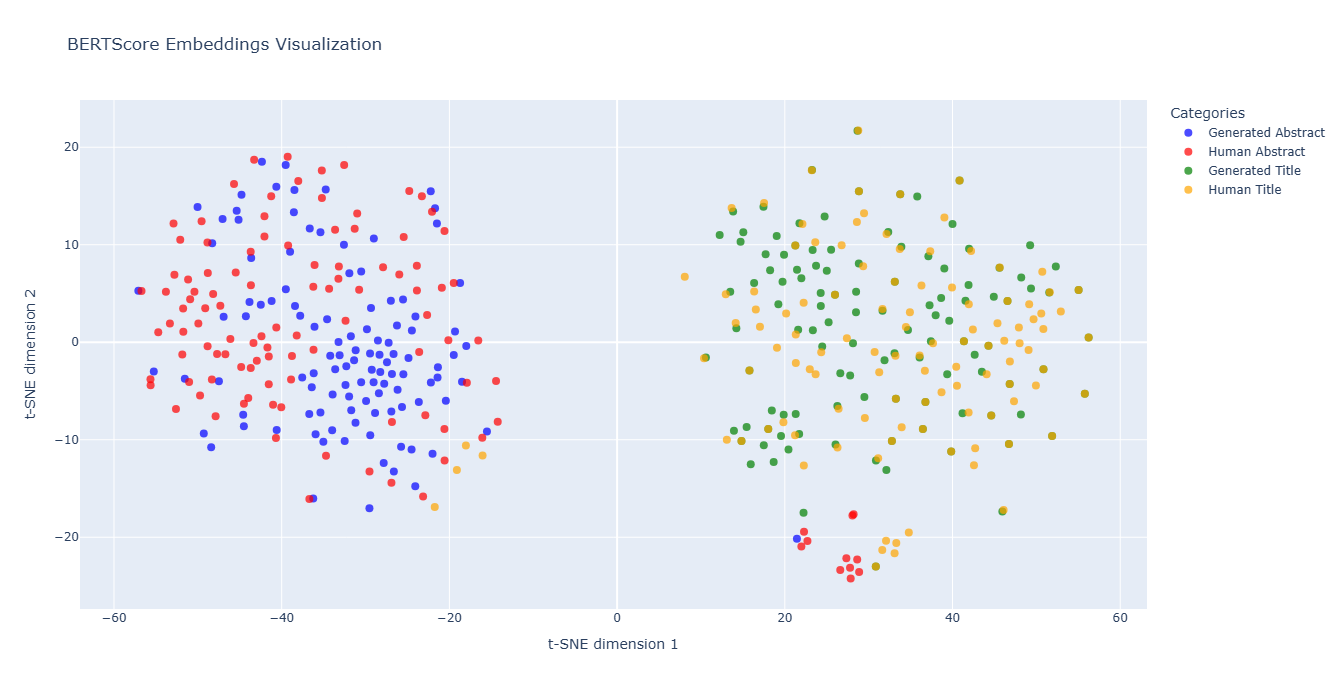}
    \caption{BERTscore for title and abstract demonstration}
    \label{fig:bertscore2d}
\end{figure}

\begin{figure}[ht]
    \centering
    \includegraphics[width=0.9\linewidth]{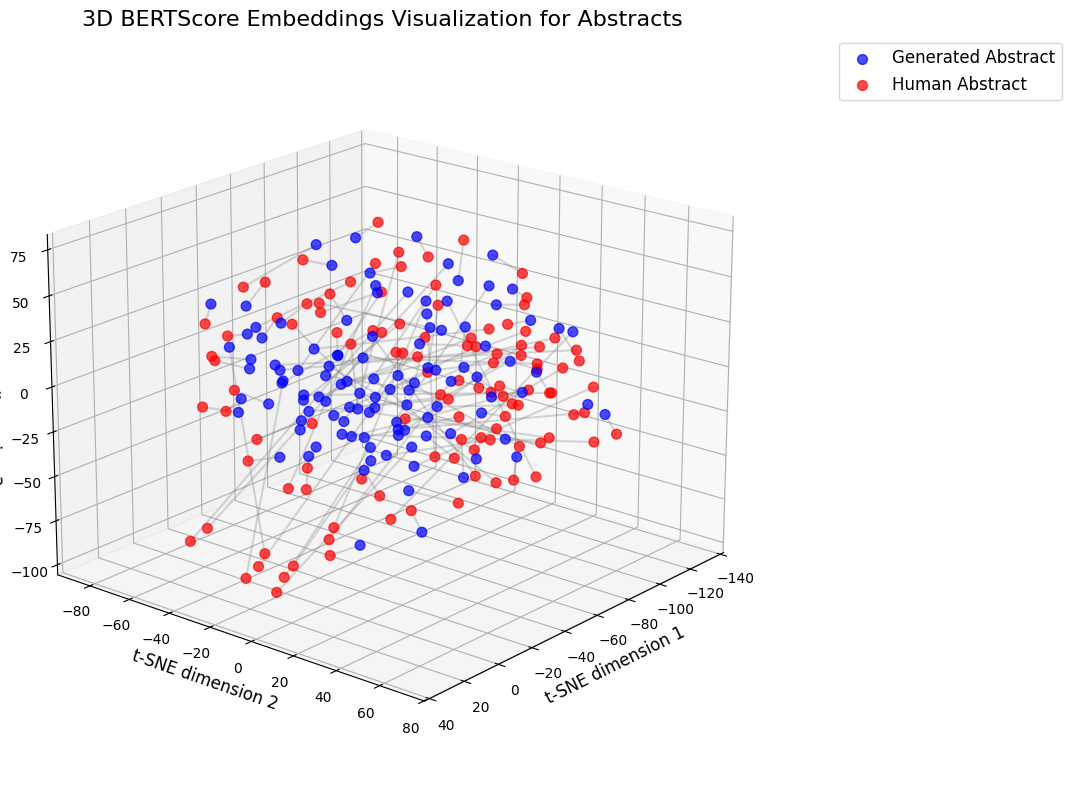}
    \caption{3D BERTscore for abstract demonstration}
    \label{fig:bertscore3d}
\end{figure}
\clearpage 

\begin{figure}[ht]
    \centering
    \includegraphics[width=0.5\textwidth]{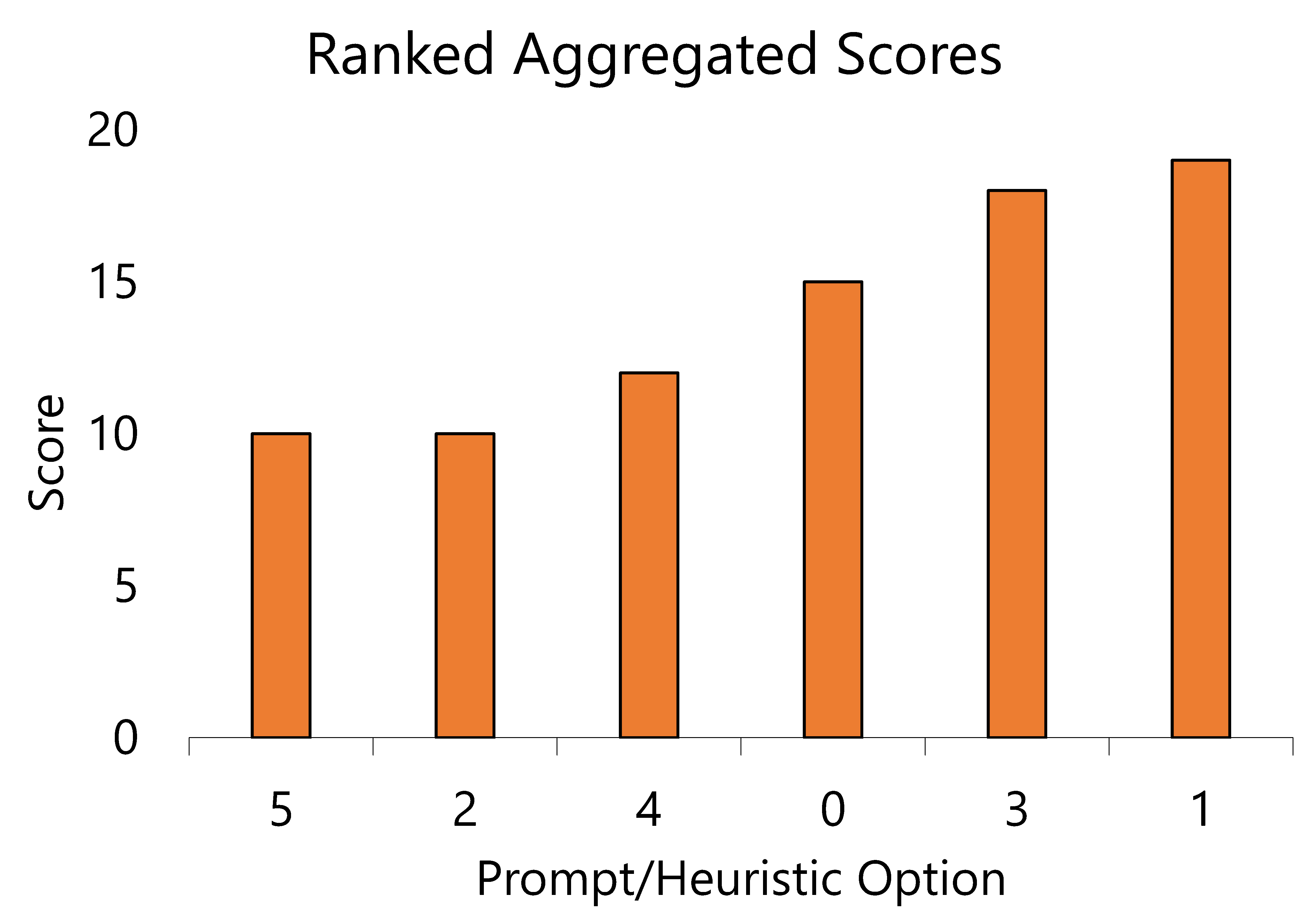}
    \caption{Bar chart of ranked aggregated scores for the six different combinations of prompts and heuristics. The scores are based on criteria including maximum BERTScore median, minimum Levenshtein median, and minimum standard deviation.}
    \label{fig:rankedscore}
\end{figure}

Our ranking algorithm evaluates heuristic variants based on their ability to generate abstracts and titles through an aggregated scoring system. This method integrates three primary criteria: the minimal median of Levenshtein distance, the maximum median of BERTScore, and the minimum standard deviation. The semantic similarity between reference and produced texts is assessed by BERTScore through contextual embeddings from BERT; a larger median indicates increased quality. The exact match accuracy is measured by the Levenshtein distance; a lower median indicates a closer alignment with the reference. The standard deviation of BERTScores is a metric that indicates consistency; values that are lower indicate more consistent performance. A total score is generated by combining the rankings of heuristics for each criterion. This comprehensive method ensures a fair evaluation of accuracy and consistency, thereby influencing our selection of heuristics for text production. The 2D and 3D visualizations of BERTScore of our generated data are demonstrated in figure \ref{fig:bertscore2d} and figure \ref{fig:rankedscore}. In the 2D BERT embedding vector space, it is evident that certain abstract pairs and title pairs are identical. The majority of the distances between abstract pairs is negligible.

\begin{table}[ht]
    \centering
    \begin{tabular}{|l|c|c|c|}
        \hline
        Prompt & Heuristic & Combination & Ranked Aggregated Scores \\
        \hline
        With Rules & 1 & 0 & 4 \\
        Without Rules & 1 & 1 & 6 \\
        \rowcolor{green!20} 
        With Rules & 2 & 2 & 1 \\
        Without Rules & 2 & 3 & 5 \\
        With Rules & 3 & 4 & 3 \\
        \rowcolor{green!20} 
        Without Rules & 3 & 5 & 1 \\
        \hline
    \end{tabular}
    \vspace{1em} 

\caption{Comparison of ranked aggregated scores for different prompt and heuristic combinations. The highlighted rows, heuristic 2 with rules and heuristic 3 without rules, achieved the best overall performance, each with a top-ranked score of 1. These combinations were selected for manual evaluation based on their superior results in automated scoring metrics.}
\end{table}

The automated evaluation showed that the following approaches had the best scores:
\begin{itemize}
    \item Heuristic 2 with 2nd prompt with rules
    \item Heuristic 3 with 2nd prompt without rules
\end{itemize}

\textbf{Manual evaluation}: 

Based on the automated evaluation results, we shortlisted the two approaches for manual evaluation. These titles and abstracts, in combination with a set generated by a human, were then subjected to manual grading by a team of eight trained cataloguers. The results were evaluated using the statistical tests. Cochran’s Q test and McNemar’s test are statistical tools that are employed to assess categorical data, particularly in situations involving related samples or repeated measurements~\cite{cochran1950,mcnemar1947}.  These tests are employed to compare the performance of human evaluation and various heuristics in our research.

McNemar’s test is often used to analyze nominal data that is paired, particularly in before-and-after studies or when comparing two related samples. It is particularly beneficial for binary results in matched pairs of subjects. In this study, we employ McNemar’s test to evaluate combination 2 with human assessment. This pairwise comparison enables us to determine whether the automated heuristic and human judgment exhibit statistically significant differences. Cochran’s Q test is an expanded version of McNemar’s test that is applicable to more than two related groups. It is implemented when three or more conditions that are related generate binary outcomes, such as “pass” or “fail.” The test assesses whether the proportions of these binary outcomes vary statistically significantly across the conditions. We apply Cochran’s Q test to determine whether the pass/fail rates of the three evaluation approaches (combination 2, combination 3, and human) display any significant variations. This test elucidates whether the approaches are consistently producing distinct results or if the variations could be the result of random chance.

In our dataset, there are numerous evaluations of the same set of objects, including both human and heuristic evaluations. Both tests are qualified by our pass/failure data. The purpose of these tests is to conduct a precise statistical comparison of the performance of various evaluation techniques. McNemar’s test enables us to focus on the comparison between two combinations and human evaluation, which may be helpful in assessing the automated approach’s performance against human judgment.

\subsection{Software and Hardware}
To verify if heuristics can effectively reduce the number of tokens needed for metadata generation, we wrote a python package warc2summary which provides a pipeline for metadata creation and evaluation. In short, this package uses WARCIO and fastwarc as the WARC file processor, and for title and abstract generation we use OpenAI API (GPT-4o) as the main model and instructor to constrain the output~\cite{warcio,bevendorff2021fastwarc,instructor2024} 

We ran the software on a Lenovo ThinkPad T14 with an 11th Gen Intel® Core™ i5-1145G7 CPU and 16GB of DDR4 RAM. The pipeline took close to two hours to process the WARC files and another two hours to generate the synthetic data from API calls. We expect faster performance in WARC file processing with a better CPU. In addition, the API call pipelines were called sequentially in order to not breach OpenAI API rate limits. With higher rate limits, it is possible to parallelise the calls for better performance. The WARC files are warc.gz files and are taken from Web Archive Singapore.
This provided us with a DataFrame containing synthetic titles and abstracts. 

We identified six different combinations of prompts and heuristics as stated above. We then used a ranked aggregation scoring method, incorporating Levenshtein distance for the title and BERTScore (using bert-based-cased embeddings), with criteria of maximum BERTScore median, minimum Levenshtein median, and minimum standard deviation for both, to shortlist two specific combinations of prompts and heuristics. Finally, a team of eight trained cataloguers performed manual grading.

\section{Results}

\subsection{ Data Reduction}
From the DataFrame, each heuristic finished within 15 seconds on 112 files. The heuristics were consolidated. Based on our heuristics, the final DataFrame should have an identical number of rows as the number of WARC files. Based on the collection of 112 WARC files we have, we obtained a 99.9\% reduction in total token count, resulting in significant cost savings compared to letting OpenAI parse the entirety of the WARC files.

\begin{figure}[ht]
   \centerline{ \includegraphics[width=0.7\textwidth]{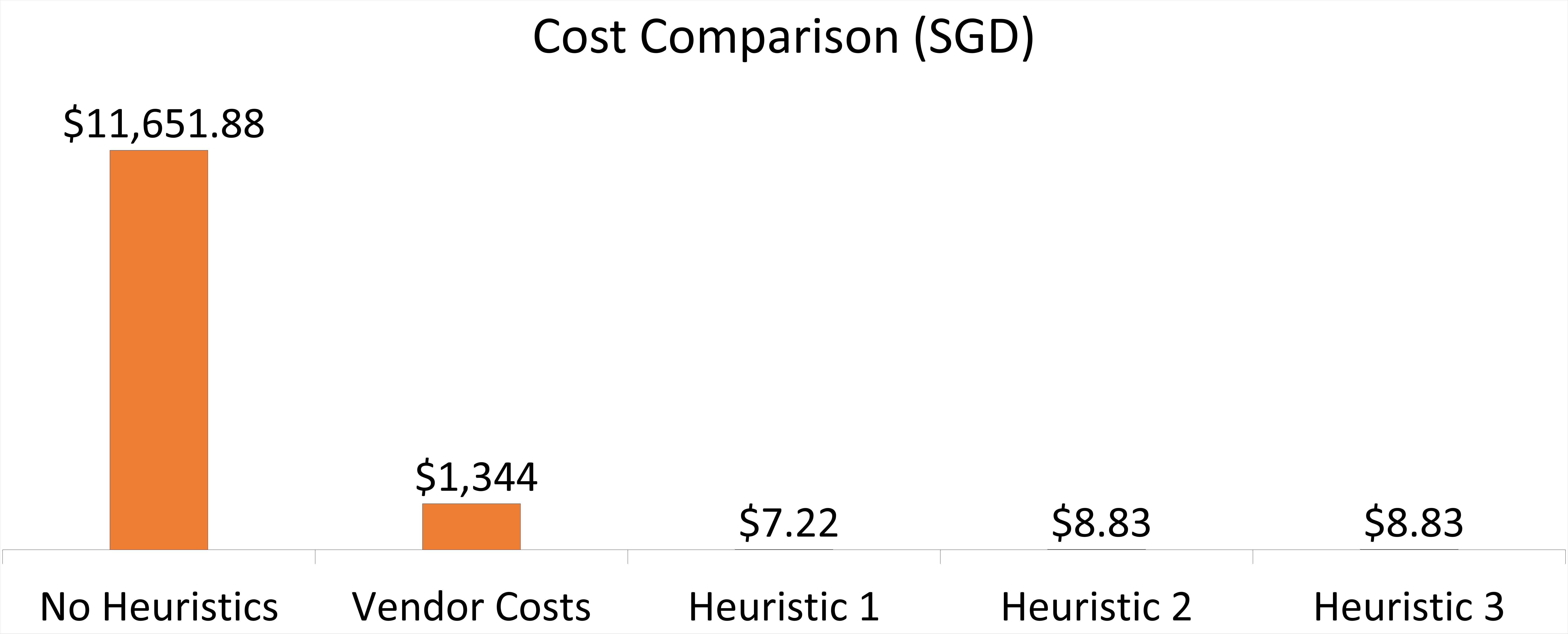}}
    \caption{Visualizing cost differences before and after applying heuristics to count tokens with tiktoken, a Byte-Pair Encoding tokenizer for GPT-4o~\cite{tiktoken}.}
    \label{fig:total costs}
\end{figure}

\subsection{Statistical Test}
We performed Cochran’s Q test by having trained cataloguers rate the titles and abstracts, without knowledge of their provenance. Based on a 5\% confidence level, we found that the synthetic titles and abstracts are statistically distinguishable from human-generated titles and abstracts. Since the p-value (0.02) is less than our significance level (\(\alpha = 0.05\)), we reject the null hypothesis, indicating a significant difference between the GPT-4o generated metadata and human crafted metadata.

\begin{figure}[ht]
    \centering
\includegraphics[width=0.6\linewidth]{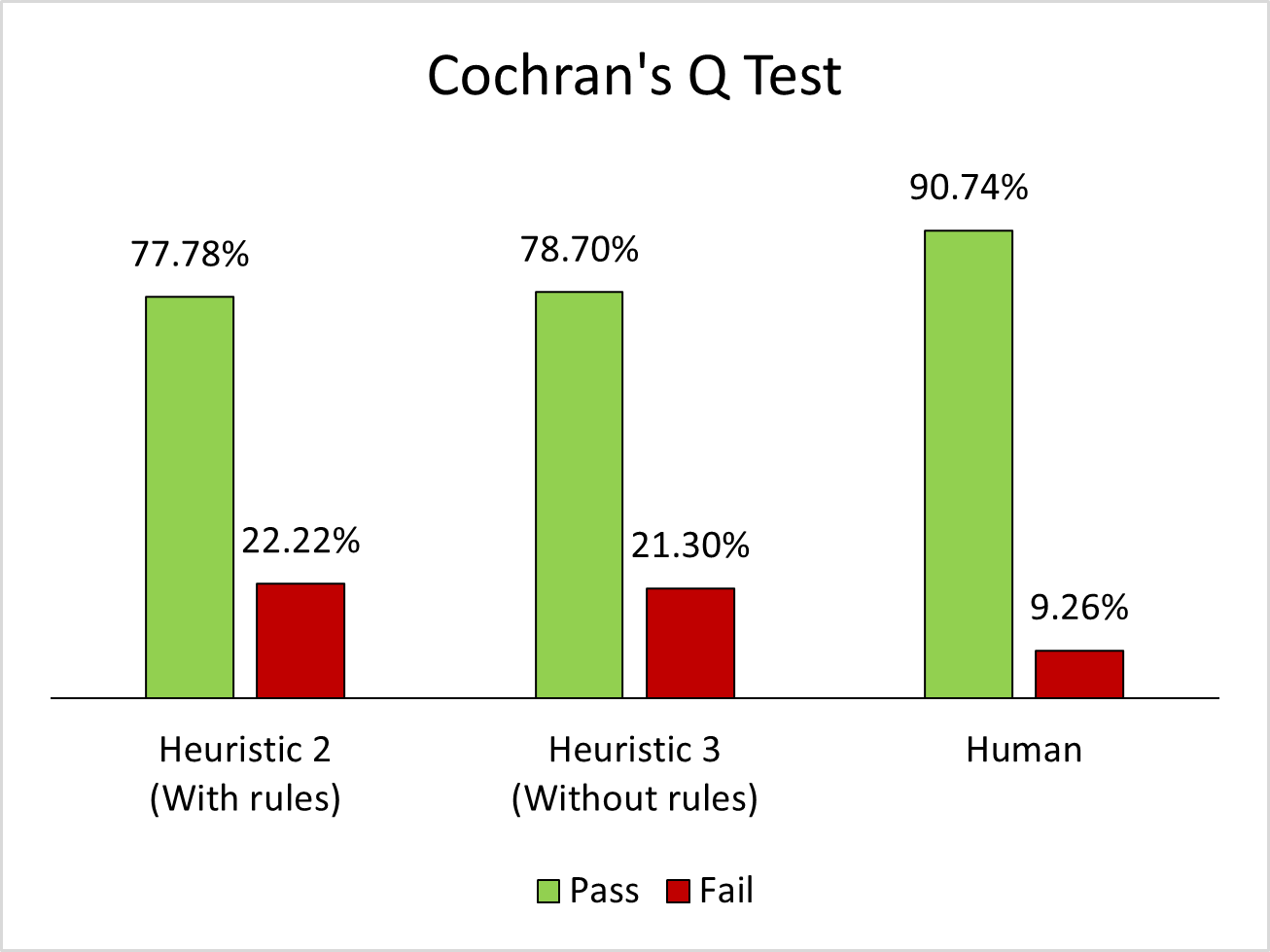}
    \caption{Results of Cochran’s Q test}
    \label{fig:Q_test}
\end{figure}

\begin{figure}[ht]
    \centering
    \includegraphics[width=0.7\linewidth]{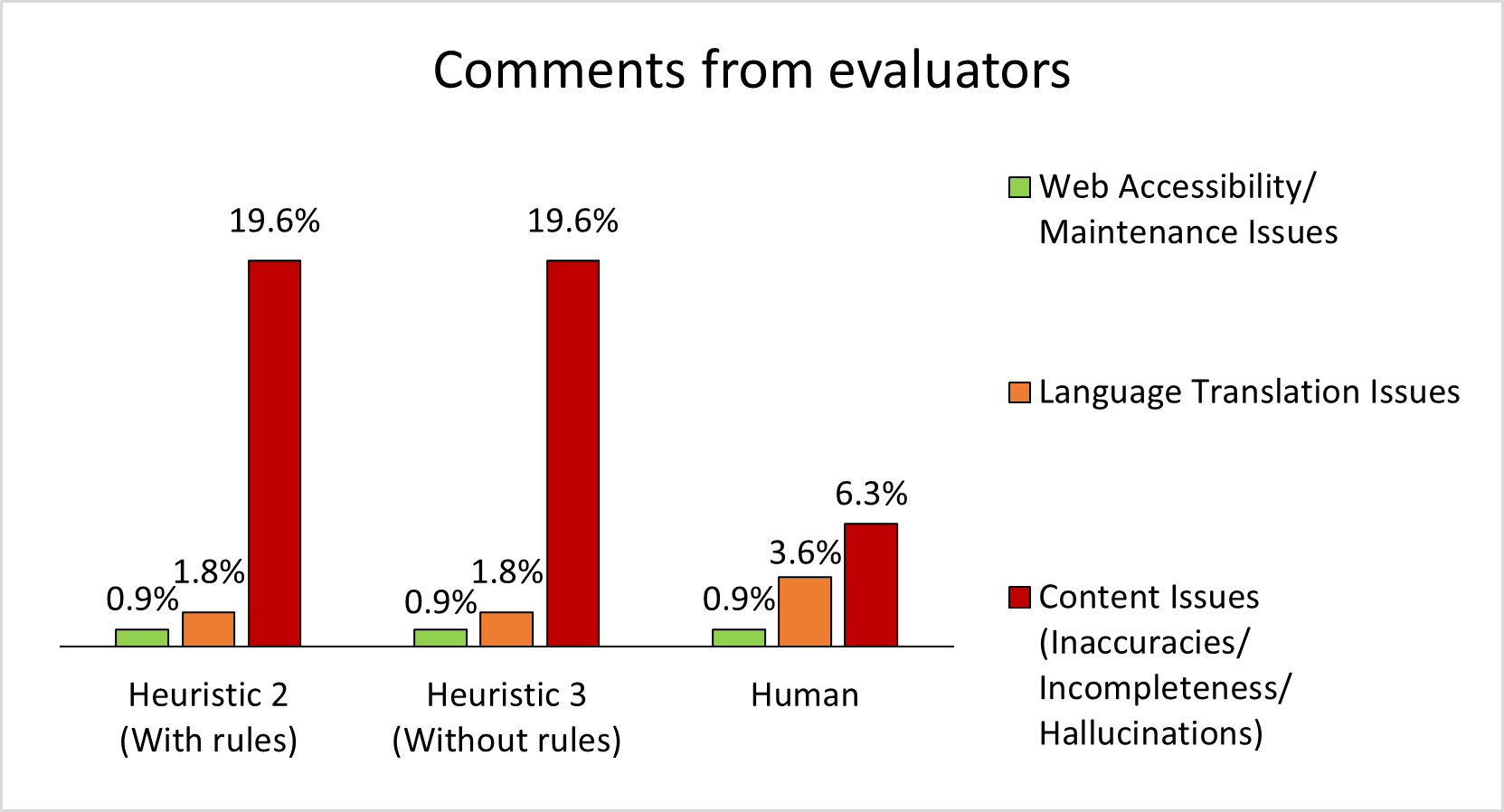}
    \caption{Issues faced from evaluator’s comments}
    \label{fig:total_costs}
\end{figure}

\section{Discussion} 
\subsection{Implications of Results}
Our study demonstrates both the potential and limitations of using large language models like GPT-4o for automated metadata generation in web archives:
\begin{enumerate}
    \item \textbf{Scalability and efficiency:} Our approach efficiently processes large volumes of web archive data, addressing a critical challenge in digital preservation. The method achieves a 99.9\% reduction in associated costs compared to full WARC processing.
    
    \item \textbf{Cost-effectiveness:} By reducing the need for manual cataloguing, our approach significantly lowers the costs associated with metadata creation, allowing institutions to reallocate resources to other critical tasks.
    
    \item \textbf{Quality comparison:} Our adaptation of the Turing test, involving eight cataloguers evaluating metadata from different sources, provided valuable insights into the current capabilities of LLMs, specifically GPT-4o in this domain. The synthetic titles and abstracts were statistically distinguishable from human curated metadata, with a p-value of 0.02, suggesting that human generated metadata still maintains an edge in quality. Furthermore, there is no significant difference between LLM based approaches, with rules and without rules.
\end{enumerate}

\subsection{Limitations and Challenges}
Our study revealed several important limitations and challenges, both from our experimental results and broader considerations in the field.

\begin{enumerate}
    \item \textbf{Content accuracy and hallucinations:} 19.6\% of LLM generated titles and abstracts had content issues such as inaccuracies, incompleteness, or hallucinations, higher than the 6.3\% observed in human-curated metadata. This highlights the known challenge of LLM hallucinations, where generated content can be factually incorrect or not grounded in the input data. Mitigating these hallucinations remains an active area of research.
    
    \item \textbf{Language translation:} Both LLM and human generated metadata faced challenges with language translation, highlighting the complexity of handling multilingual content in web archives.
    
    \item \textbf{Legal considerations:} The generation and use of metadata from copyrighted web content raises complex legal questions. Navigating fair use and copyright in the context of web archiving and LLM generated metadata requires careful consideration to avoid potential legal complications~\cite{zakir2024navigating}.
    
    \item \textbf{Privacy and dependency concerns:} The automated nature of LLMs for metadata generation poses challenges in honoring individual requests for content removal or anonymization, which are crucial aspects of privacy rights. Additionally, our reliance on proprietary, closed-source LLMs like GPT-4o introduces dependencies that may limit flexibility and control. LLMs can potentially leak training data, raising concerns about privacy and data protection~\cite{carlini2021extracting}.  The potential for these models to use web content for further training underscores the need for transparency and alternatives.
    
    \item \textbf{Quality of input data:} The adage "garbage in, garbage out" is especially relevant in our context.  The quality of LLM generated metadata is directly dependent on the quality of the input data from web archives. Ensuring that our web archive data is accurate, diverse, well labelled, and free from noise is crucial for generating reliable and unbiased metadata~\cite{mellin1957, babbage1864}.
    
    \item \textbf{Evaluation criteria:} Ensuring that LLM generated content accurately represents the website, remains relevant to the topic, and avoids subjective interpretations or biased language remains a significant challenge, as evidenced by our experimental results.
\end{enumerate}

\section{Conclusion}

This study introduces a novel approach to generating metadata for web archives using GPT-4o, combining data reduction heuristics with evaluation methods inspired by the Turing test. Our findings highlight the efficiency and scalability of LLM generated metadata, though it still falls short in quality and accuracy compared to human-curated metadata. GPT-4o produced more inaccuracies and hallucinations but is cost effective and scalable for WARC file metadata creation. This suggests LLMs are best as assistive tools for human cataloguers rather than replacements. Future efforts could focus on changing the prompts to fit the cataloguers’ standards, more aggressive heuristics to filter out promotional website content, developing strategies to reduce and identify hallucinations and using smaller language models to circumvent privacy concerns.

Our study marks a significant step towards leveraging AI in web archiving, offering valuable insights into its current capabilities and limitations. Addressing the identified challenges will help us work towards a future where AI enhances the preservation and accessibility of digital heritage, maintaining high standards crucial for the utility of web archives. This research sets a roadmap for future improvements, aiming to bridge the quality gap between AI and human curated metadata.
Our study marks a significant step towards leveraging AI in web archiving, offering valuable insights into its current capabilities and limitations. Addressing the identified challenges will help us work towards a future where AI enhances the preservation and accessibility of digital heritage, maintaining high standards crucial for the utility of web archives. This research sets a roadmap for future improvements, aiming to bridge the quality gap between AI and human curated metadata.

\section{Acknowledgments}We would like to extend our heartfelt gratitude to the following cataloguers from the National Library Board for their invaluable contributions to this project:
\begin{itemize}
    \item Ann Cheah
    \item Chang Siew Fen
    \item Chor Swee Chin
    \item Munifah Shaik Mohsin Bamadhaj
    \item Ng Wee Lay
    \item Rohaya Bte Yacob
    \item Tan Minli Mindy
\end{itemize}

\bibliographystyle{unsrtnat}
\bibliography{references}

\end{document}